# ASYMMETRIC POLYCYCLIC AROMATIC HYDROCARBON AS A CAPABLE SOURCE OF ASTRONOMICALLY OBSERVED INTERSTELLAR INFRARED SPECTRUM

NORIO OTA

Graduate School of Pure and Applied Sciences, University of Tsukuba,
1-1-1 Tenoudai Tsukuba-city 305-8571, Japan; n-otajitaku@nifty.com

In order to find out capable molecular source of astronomically well observed infrared (IR) spectrum, asymmetric molecular configuration polycyclic aromatic hydrocarbon (PAH) was analyzed by the density functional theory (DFT) analysis. Starting molecules were benzene $C_6H_6$, naphthalene $C_{10}H_8$ and 1H-phenalene $C_{13}H_9$. In interstellar space, those molecules will be attacked by high energy photon and proton, which may bring cationic molecules as like $C_6H_6^{n+}$ (n=0~3 in calculation), $C_{10}H_8^{n+}$, and $C_{13}H_9^{n+}$, also CH lacked molecules $C_5H_5^{n+}$, $C_9H_7^{n+}$, and $C_{12}H_8^{n+}$. IR spectra of those molecules were analyzed based on DFT based Gaussian program. Results suggested that symmetrical configuration molecules as like benzene, naphthalene, 1H-phenalene and those cation ( +, 2+, and 3+) show little resemblance with observed IR. Contrast to such symmetrical molecules, several cases among cationic and asymmetric configuration molecules show fairly good IR tendency. One typical example was $C_{12}H_8^{3+}$, of which calculated harmonic IR wavelength were 3.2, 6.3, 7.5, 7.8, 8.7, 11.3, and 12.8µm, which correspond well to astronomically observed wavelength of 3.3, 6.2, 7.6, 7.8, 8.6, 11.2, and 12.7µm. It was amazing agreement. Also, some cases like $C_5H_5^+$, $C_9H_7^+$, $C_9H_7^{2+}$, $C_9H_7^{3+}$ and $C_{12}H_8^{2+}$ show fairly good coincidence. Such results suggest that asymmetric and cationic PAH may be capable source of interstellar dust.

Key words: astrochemistry · infrared: numerical · molecular data: PAH – benzene, naphthalene, phenalene:

## 1, INTRODUCTION

It is well known that interstellar polycyclic aromatic hydrocarbon (PAH) shows ubiquitous specific infrared (IR) spectrum from 3 to 20µm (Boersma et al. 2014). Recently, Tielens pointed out in his review (Tielens 2013) that void induced polycyclic aromatic hydrocarbons (PAH's) may be one candidate. According to such previous expectations, we tried to test theoretically on one typical example of void coronene $C_{23}H_{12}^{++}$ (Ota 2014, 2015a, 2015b). Theoretical calculation was done based on the density functional theory using Gaussian09 package (Frisch et al. 2009, 1984). Result was amazing that this single molecule could almost reproduce a similar IR spectrum with astronomically well observed one. The current central concept to understand observed astronomical spectra is the decomposition method from the data base of many PAH's experimental and theoretical analysis (Boersma et al. 2013, 2014). Question is why previous calculation of single molecule $C_{23}H_{12}^{++}$ could almost reproduce observed spectra. In order to explain such behavior, naphthalene $C_{10}H_8$ and its void induced cation $C_9H_7^{n+}$ was analyzed in a previous note (Ota 2015c). In interstellar space, especially new star born area, high energy proton and photon attack PAH molecules to bring void induced PAH and its cation. Symmetrical configuration of naphthalene and its cation has no electric dipole moment and show little similarity with observed IR. Whereas, it's void induced carbon pentagon-hexagon molecule $C_9H_7^{n+}$ has intrinsic electric dipole moment by its asymmetrical geometry and may bring fairly good IR behavior. Here, we like to continue such trial by enhancing cases starting from benzene $C_6H_6$ (single carbon ring) to naphthalene $C_{10}H_8$ (two rings), and 1H-phenalene $C_{13}H_9$ (three rings) and will compare their void induced cation $C_5H_5^{n+}$, $C_9H_7^{n+}$, and $C_{12}H_8^{n+}$. Again, it will be demonstrated that asymmetrical molecular configuration with carbon pentagon ring bring specific IR behavior similar with observed one. This paper suggests that asymmetric cationic PAH's are capable candidates in interstellar space.

## 2, MODEL MOLECULES

As a typical example of model molecule, as illustrated in Figure 1(a), ionization and void creation of benzene $C_6H_6$ was illustrated. In interstellar space, especially in new star born area, high energy proton and high energy photon may attack benzene. Proton attacks one particular carbon site of $C_6H_6$ and creates a single void. After that, recombination of molecule may bring to $C_5H_5$, which is shown in Figure 1(b). At the same time, high energy photon attacks such molecules to ionize deeply such as cation $C_6H_6^+$, $C_6H_6^{++}$ as $(C_6H_6)^{n+}$ (n=0 to 3 in calculation), also creates void induced (CH lacked) cation $C_5H_5^{n+}$. For every ionized state, IR spectrum calculation should be done after atom position optimization. As shown in Table 1, we enhanced test molecules from benzene $C_6H_6$ (single carbon ring) to naphthalene $C_{10}H_8$ (two rings), 1H-phenalene $C_{13}H_9$ (three rings) and their void induced molecules $C_5H_5^{n+}$, $C_9H_7^{n+}$, and $C_{12}H_8^{n+}$ (n=0 to 3).

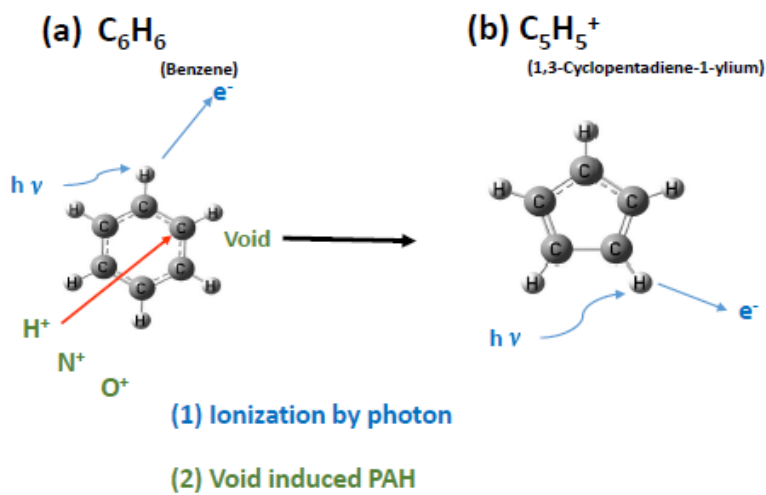

**Figure 1** Modeling a creation of cation and void in PAH. In interstellar space, high energy photon attacks on PAH. For example, benzene $C_6H_6$ may turn to cationic molecules like $C_6H_6^+$, $C_6H_6^{2+}$ etc. Also, high energy proton attacks carbon site and brings void. By a molecular recombination process, new void induced PAH will be created as like $C_5H_5$ in Figure 1(b), which is also ionized as like $C_5H_5^+$.

**Table 1** Model molecules from single carbon ring to three rings. Void induced molecules are illustrated in lower line. Every molecule was ionized by high energy photon irradiation as PAH $^{n+}$ (charge number n=0 to 3 in calculation).

|  | 1-ring | 2-rings | 3-rings |
|---|---|---|---|
| PAH | $C_6H_6$ | $C_{10}H_8$ | $C_{13}H_9$ |
| Void induced PAH | $C_5H_5$ | $C_9H_7$ | $C_{12}H_8$ |

3, CALCULATION METHOD

We have to obtain total energy, optimized atom configuration, and infrared vibrational mode frequency and strength depend on a given initial atomic configuration, charge and spin state Sz. Density functional theory (DFT) with unrestricted B3LYP functional (Becke 1993) was applied utilizing Gaussian09 package (Frisch et al. 2009, 1984) employing an atomic orbital 6-31G basis set. The first step calculation is to obtain the self-consistent energy, optimized atomic configuration and spin density. Required convergence on the root mean square density matrix was less than $10^{-8}$ within 128 cycles. Based on such optimized results, harmonic vibrational frequency and strength was calculated. Vibration strength is obtained as molar absorption coefficient ε (km/mol.). Comparing DFT harmonic wavenumber $N_{DFT}$ (cm$^{-1}$) with experimental data, a single scale factor 0.965 was used (Ota 2015b). For the anharmonic correction, a redshift of 15cm$^{-1}$ was applied (Ricca et al. 2012).

Corrected wave number N is obtained simply by N (cm$^{-1}$) = $N_{DFT}$ (cm$^{-1}$) x 0.965 – 15 (cm$^{-1}$).

Also, wavelength λ is obtained by λ (μm) = 10000/N(cm$^{-1}$).

4, BENZENE $C_6H_6^{n+}$ AND VOID INDUCED $C_5H_5^{n+}$

Calculated IR spectra of benzene and its cation was illustrated in Figure 2. Vertical value show integrated absorption coefficient ε (km/mol.), whereas horizontal line shows wavelength λ (μm). Blue curve was calculated Lorentzian type distribution with the full width at half-maximum (FWHM) of 15 cm$^{-1}$ based on the harmonic intensity. In every case, electric dipole moment D was just zero by its symmetrical molecular configuration. In case of neutral $C_6H_6$, major IR peaks were 3.2, 6.8, 9.8, and 15.0μm. In mono-cation $C_6H_6^+$, they were 3.2, 6.7, 7.0, 7.4, 10.8, and 15.1μm. Except 3.2μm, those peaks are not coincident with astronomically well observed wavelength 6.2, 7.6, 7.8, 8.6, 11.2, 12.7, 13.5, 14.3μm in an accuracy of +/- 0.2μm. Di-cation $C_6H_6^{2+}$ case, we can see coincidence only at 3.2 and 13.5μm. Calculation of tri-cation $C_6H_6^{3+}$ could not terminated due to serious molecular deformation.

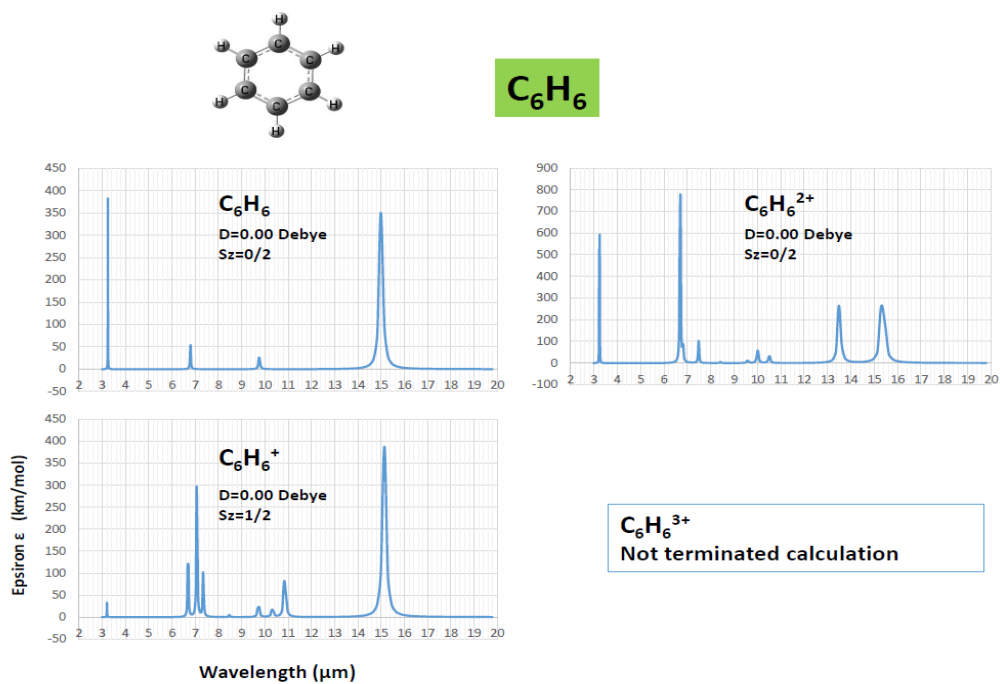

Figure 2, Calculated IR spectra of ionized benzene $C_6H_6^{n+}$. Typical strong peaks were observed at 3.2, 7.0, 15.0μm. Except 3.2μm, we could not see any good coincidence with astronomically well observed peaks as like 6.2, 7.6, 7.8, 8.6, 11.2μm etc.

IR spectra of void induced group $C_5H_5^{n+}$ were shown in Figure 3. Electric dipole moment D of those group were none-zero intrinsic values depend on their asymmetric molecular configuration. Neutral $C_5H_5$ is radical with spin parameter Sz=1/2, which may bring unstable short life time on conventional condition. However, in interstellar circumstance, it may have longer life time due to extra few interaction chances with other molecules. Calculated peaks of $C_5H_5$ were 3.2, 7.0, 7.4, and 15.1μm, which are not related with observed spectrum except 3.2μm. In case of mono-cation $C_5H_5^+$, wavelength of 3.2, 6.5, and 7.7μm were fairly good coincidence, but not good at 8.1, 10.7, and 14.8μm. In Figure 4, calculated IR of mono-cation $C_5H_5^+$ was compared with observed emission spectra from four astronomical sources by Boersma et al. (Boersma 2009). Fundamental harmonic wavelength λ (μm) and absorption coefficient ε (km/mol.) and their vibrational mode were summarized in Table 2. Observed wavelength are marked by green in the left column. Blue marked wavelength show good coincidence with observed one within +/- 0.2μm, while red one are out of this range. Vibrational mode analysis was also done and noted in right column. C-H stretching at 3.2μm and C-H in-plane bending at 7.7μm are typical modes of PAH. Calculated peak of 6.5μm is 0.3μm longer than observed one, which is marked by red. Also, calculated 10.6 and 10.8μm are red marked, which are related to observed 11.2μm wavelength. Total calculated IR behavior of $C_5H_5^+$ almost traces observed spectrum, but shows no good quantitative coincidence. In case of di-cation $C_5H_5^{2+}$, we could find coincidence at 3.2 and 7.7μm, other peaks at 7.1, 7.2, 11.9, and 15,4μm were not. In case of tri-cation $C_5H_5^{3+}$, calculation did not terminated by serious molecular deformation.

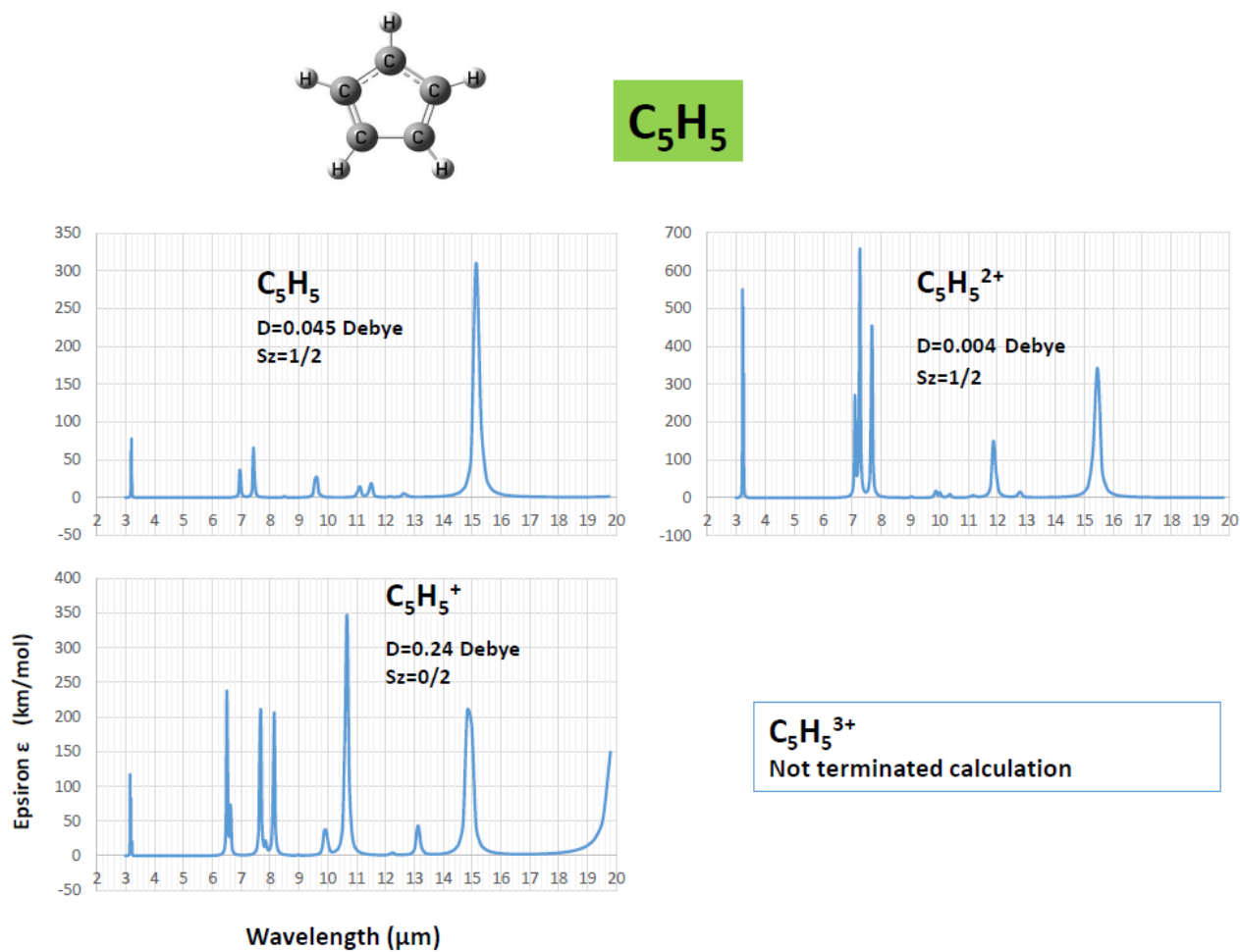

**Figure 3,** Calculated IR spectra of $C_5H_5^{n+}$. Mono-cation $C_5H_5^+$ show fairly good behavior at 3.3, 7~8, and 11μm bands, but no good quantitative agreement with observed one.

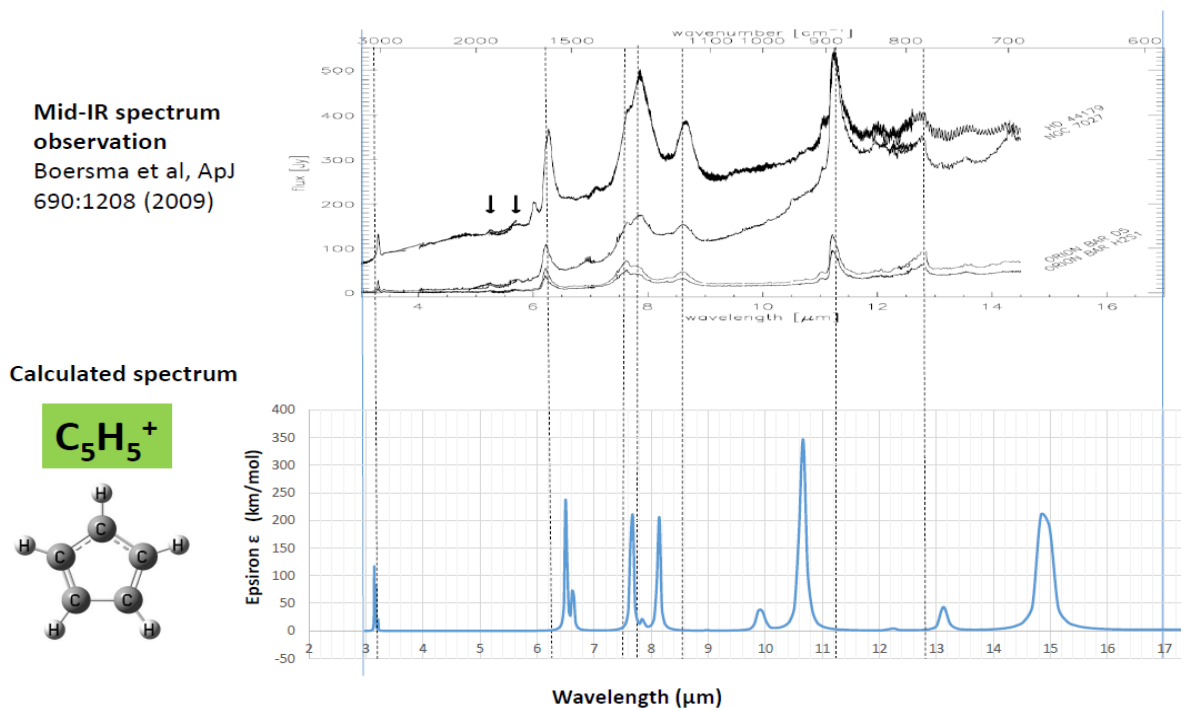

Figure 4, Calculated IR of $C_5H_5^+$ (lower figure) was compared with observed emission spectra (upper one) from four astronomical sources edited by Boersma et al. (Boersma 2009). Calculated IR of $C_5H_5^+$ show almost similar tendency, but no good quantitative agreement with observed lines shown by broken lines.

Table 2, Harmonic IR spectrum and vibrational mode analysis of mono-cation C5H5+. Blue marked numbers are good agreement with observed wavelength (green number) within +/- 0.2µm, while red numbers are out of range.

| Observed wavelength (micron) | Mode number | Wavelength (micron) | Epsiron (km/mol) | Vibrational mode |
|---|---|---|---|---|
| | 1 | 33.5 | 32 | C-C and C-H out of plane bending |
| | 2 | 29.8 | 0 | |
| | 3 | 20.1 | 254 | C-C in plane stretching |
| 14.3 | 4 | 14.9 | 101 | C-H out of plane bending |
| 13.5 | 5 | 13.1 | 13 | C-C in plane stretching, C-H inplane bending |
| 12.7 | 6 | 12.9 | 0 | |
| | 7 | 12.2 | 1 | |
| 11.2 | 8 | 10.8 | 13 | C-H out of plane bending |
| | 9 | 10.6 | 124 | C-C in plane stretching, C-H inplane bending |
| | 10 | 10.5 | 0 | |
| | 11 | 10.0 | 10 | |
| | 12 | 9.9 | 10 | |
| | 13 | 9.3 | 0 | |
| | 14 | 9.0 | 0 | |
| 8.6 | 15 | 8.1 | 70 | C-C in plane stretching, C-H in plane bending |
| 7.8 | 16 | 7.9 | 6 | |
| 7.6 | 17 | 7.7 | 93 | C-H in plane bending |
| | 18 | 6.6 | 32 | |
| 6.2 | 19 | 6.5 | 88 | C-C in plane stretching, |
| | 20 | 3.2 | 6 | |
| | 21 | 3.2 | 1 | |
| 3.3 | 22 | 3.2 | 23 | C-H stretching |
| | 23 | 3.1 | 34 | |
| | 24 | 3.1 | 1 | |

5, NAPHTALENE $C_{10}H_8{}^{n+}$ AND VOID INDUCED $C_9H_7{}^{n+}$

There are no electric dipole moment on naphthalene and its ionized molecules $C_{10}H_8{}^{n+}$ because of its symmetrical molecular configuration. As shown in Figure 5, pure naphthalene $C_{10}H_8$ show only two major IR peaks at 3.2 and 12.9μm, which are related to 3.3 and 12.7μm observed one. There are no complex and detailed IR structure resemble to observe one. Mono-cation $C_{10}H_8{}^+$ has three peaks at 6.7, 8.3, and 13.2μm, which do not match with observed values. Also in case of di-cation $C_{10}H_8{}^{2+}$ and tri-cation $C_{10}H_8{}^{3+}$, we could not recognize any good similarity with observed spectrum.

Whereas, void induced (CH lacked) molecule $C_9H_7{}^{n+}$ show preferable behavior as shown in Figure 6. Especially in case of mono-cation $C_9H_7{}^+$, peaks at 3.2, 6.2, 7.7, 8.6, 12.8, and 13.5μm correspond well to observe one. Direct comparison of calculated spectrum with observed one was shown in Figure 7. In Table 2, harmonic wavelength and vibrational modes were summarized. There are typical in-plane C-C stretching mode at 6.2μm, C-H in-plane bending at 8.6μm, C-H out of plane bending at 12.8, and 13.5μm. However, there are several non-fit wavelength, that is, different with observed one, such as 6.7, 7.3, 8.2μm. Also, in case of di-cation $C_9H_7{}^{2+}$ we can see corresponding good wavelength at 3.2, 7.6, 8.5, and 12.7μm, whereas no good at 7.0, 10.5μm. Tri-cation $C_9H_7{}^{3+}$ has good coincidence at 3.2, 7.6, 8.4, and 11.2μm, but no good at 7.0 and 12.2μm.

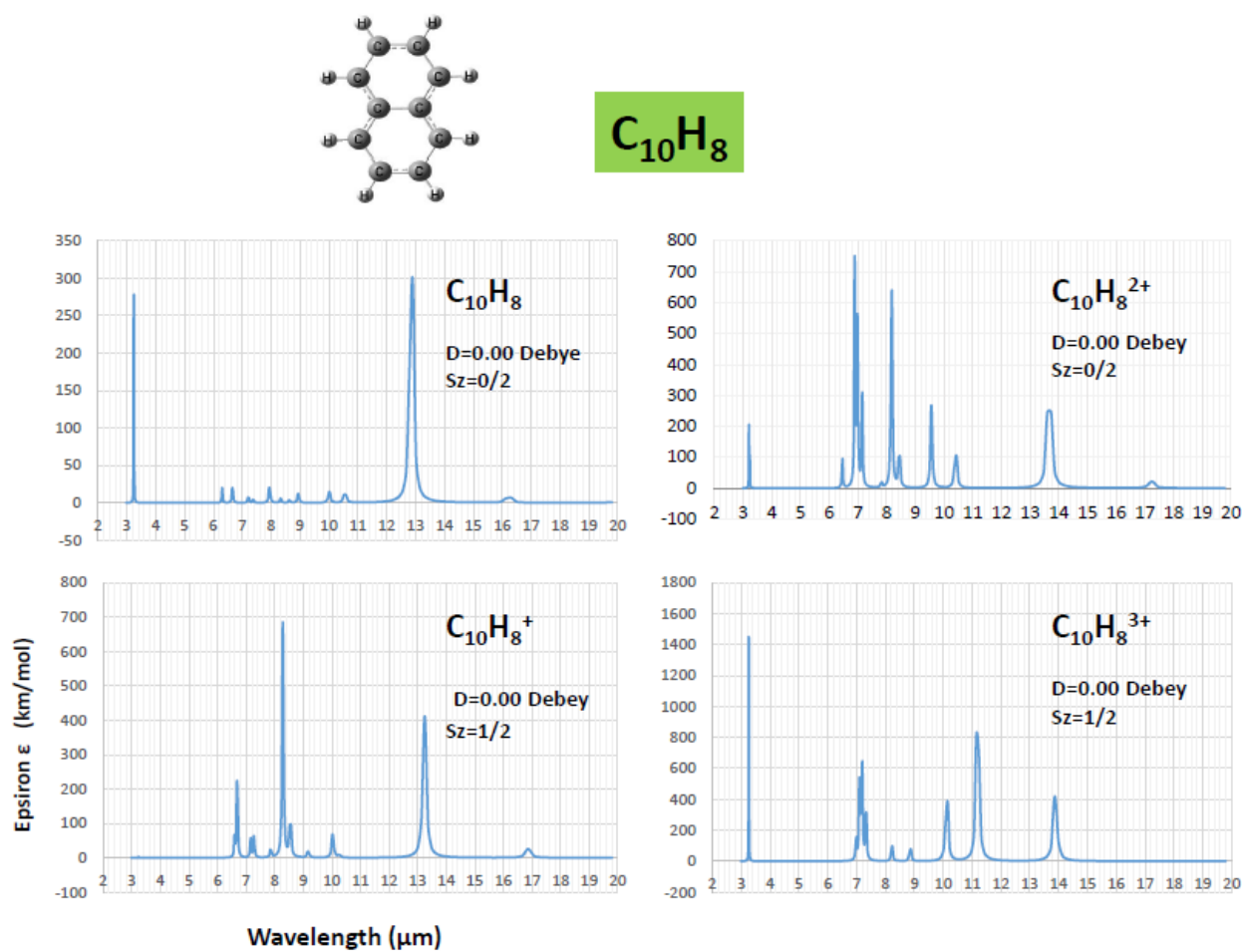

Figure 5, Calculated IR spectra of $C_{10}H_8{}^{n+}$. In every case, electric dipole moment D is zero and there are not any good similarity with astronomically observed spectrum.

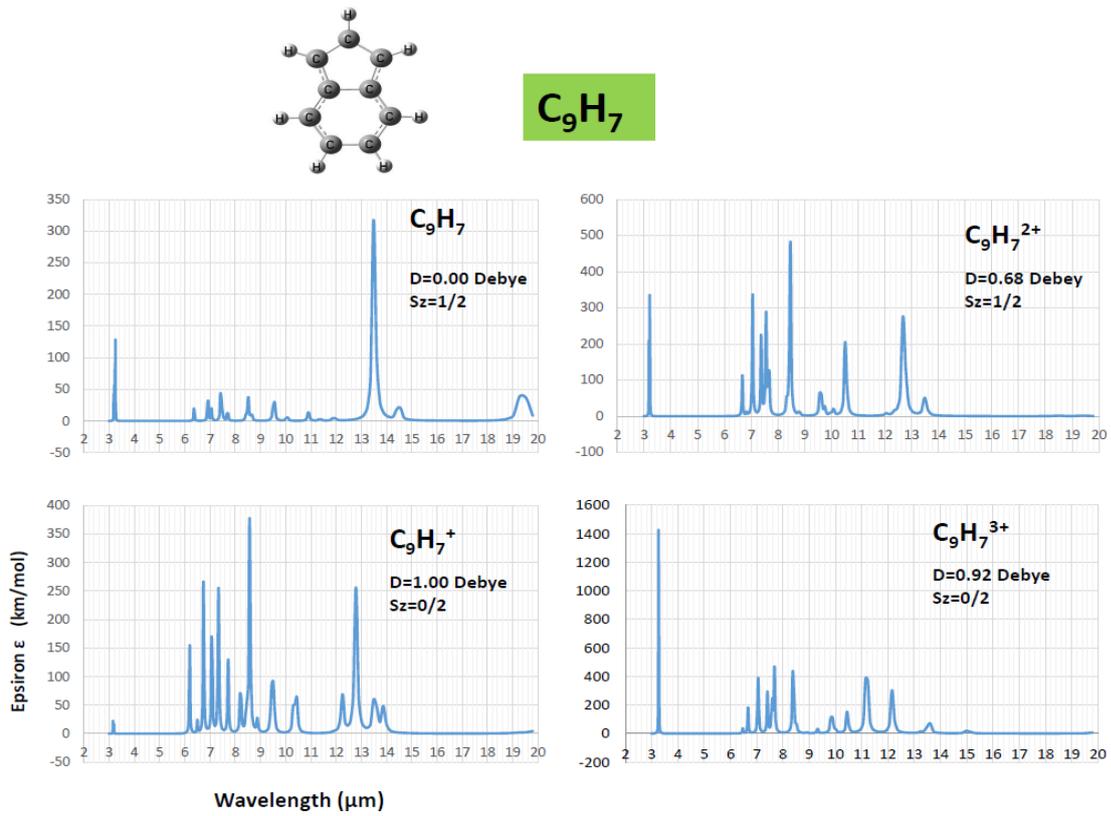

Figure 6, Calculated IR spectra of void induced (CH lacked) naphtalene $C_9H_7^{n+}$.

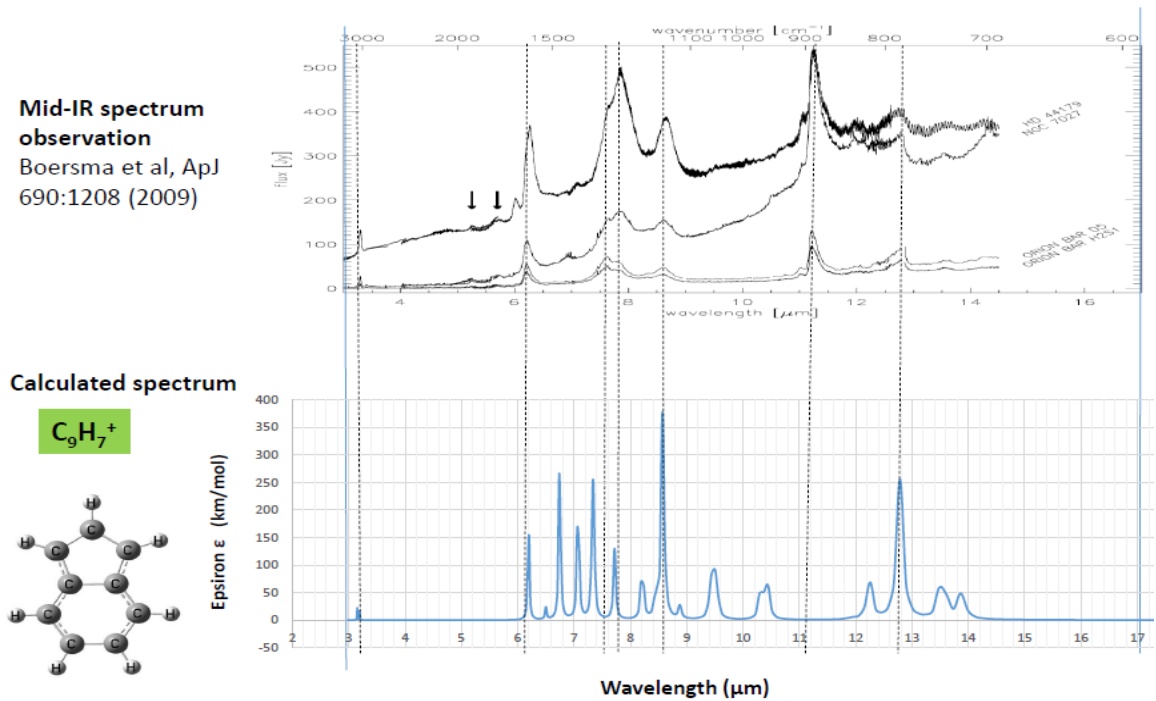

Figure 7, Comparison of calculated IR peaks of $C_9H_7^+$ with astronomically observed spectra. We can see fairly good behavior at 3.2, 6.2, 7.7, 8.6, 12.8, and 13.5μm.

Table 3, Harmonic infrared wavelength and vibrational mode analysis of $C_9H_7^+$.

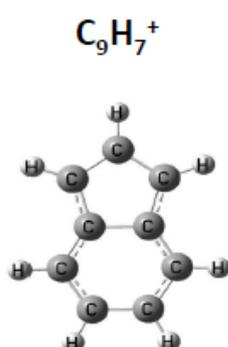

### IR spectrum of $C_9H_7^+$

| Observed wavelength (micron) | Calculated harmonic IR spectrum | | | |
|---|---|---|---|---|
| | Mode number | Wavelength (micron) | Epsiron (km/mol) | Vibrational mode |
| | 1 | 61.4 | 6 | |
| | 2 | 47.5 | 0 | |
| | 3 | 29.5 | 10 | |
| | 4 | 28.5 | 10 | |
| | 5 | 22.2 | 32 | C-C and C-H out of plane bending |
| | 6 | 20.9 | 0 | |
| | 7 | 20.4 | 17 | |
| | 8 | 19.2 | 0 | |
| 14.3 | 9 | 14.3 | 0 | |
| | 10 | 13.9 | 13 | |
| 13.5 | 11 | 13.5 | 25 | C-H out of plane bending |
| 12.7 | 12 | 12.8 | 74 | C-H out of plane bending |
| | 13 | 12.2 | 20 | |
| | 14 | 11.9 | 1 | |
| 11.3 | 15 | 11.4 | 0 | |
| | 16 | 10.5 | 4 | |
| | 17 | 10.4 | 20 | |
| | 18 | 10.3 | 14 | |
| | 19 | 10.3 | 0 | |
| | 20 | 10.3 | 0 | |
| | 21 | 10.0 | 0 | |
| | 22 | 9.5 | 27 | |
| | 23 | 9.4 | 22 | |
| | 24 | 8.9 | 7 | |
| 8.6 | 25 | 8.6 | 115 | C-H in plane bending, C-C stretching |
| | 26 | 8.4 | 14 | |
| | 27 | 8.2 | 32 | |
| 7.8 | 28 | 7.7 | 37 | |
| 7.6 | 29 | 7.3 | 77 | C-H in plane bending |
| | 30 | 7.3 | 14 | |
| | 31 | 7.1 | 74 | C-H in plane bending |
| | 32 | 6.9 | 0 | |
| | 33 | 6.7 | 95 | C-C in plane stretching |
| | 34 | 6.5 | 6 | |
| 6.2 | 35 | 6.2 | 58 | C-C in plane stretching |
| | 36 | 3.2 | 0 | |
| | 37 | 3.2 | 0 | |
| | 38 | 3.2 | 0 | |
| | 39 | 3.2 | 0 | |
| | 40 | 3.2 | 0 | |
| 3.3 | 41 | 3.2 | 6 | C-H stretching at pentagon |
| | 42 | 3.1 | 7 | C-H stretching at pentagon |

### 5, 1H-PHENALENE $C_{13}H_9{}^{n+}$ AND VOID INDUCED $C_{12}H_8{}^{n+}$

Typical compact three rings molecule is 1H-phenalene $C_{13}H_9$, which has Sz value of 1/2 as radical PAH. In Figure 8, we can see calculated peaks at 3.2 and 13.4μm related to observed 3.3 and 13.5μm wavelength, but not fit at 12.2μm. Number of major peaks are few and do not trace observed spectra. Mono-cation $C_{13}H_9{}^+$ show peaks at 6.3, 7.3, 8.0, 11.9, and 13.6μm. Among them, 6.3 and 13.6μm are related to observed wavelength, but others are not. Both di-cation $C_{13}H_9{}^+$ and tri-cation $C_{13}H_9{}^{2+}$ show no good coincidence with observed one.

Void induced cases $C_{12}H_8{}^{n+}$ were illustrated in Figure 9. Neutral $C_{12}H_8$ has poor coincidence with observed spectrum. Mono-cation $C_{12}H_8{}^+$ show fairly good agreement at 3.2, 6.4, 7.7, 8.0, and 8.5μm, whereas no good at 7.1, 8.0, 9.2, 10.0, 12.0, and 13.8μm. Also, di-cation $C_{12}H_8{}^{2+}$ show good at 3.2, 6.2, 7.8, 8.5, 11.4 and 14.4, μm, but no good at 6.9, 7.1, 8.3, 9.2, 10.0, and 12.3μm. Remarkable result was obtained in tri-cation $C_{12}H_8{}^{3+}$. Peaks at 3.2, 6.4, 7.5, 7.8, 8.0 and 11.3μm are equivalent to well observed 3.3, 6.2, 7.6, 7.8, and 11.2μm, which result is again confirmed by direct comparison with observed spectra in Figure 10. We can see good coincidence at large peaks, but there remains non-fit small peaks at 10.6 and 11.6μm. There are analyzed 54 harmonic wavelength as summarized in Table 4. We can recognize C-H stretching at 3.2~3.3μm, C-C stretching at 6.3μm, C-H in-plane bending at 7.5μm, C-C stretching and C-H in-plane bending at 7.8, 8.7μm, and 11.3μm.

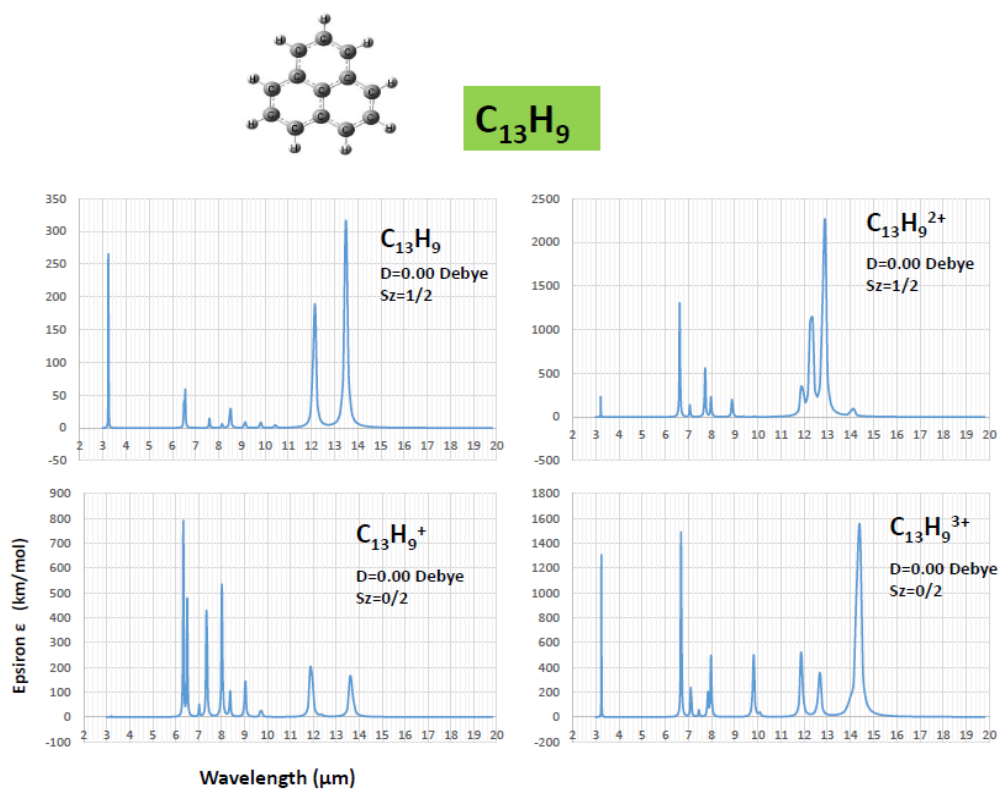

Figure 8, Calculated IR spectra of 1H-phenalene $C_{13}H_9^{n+}$.

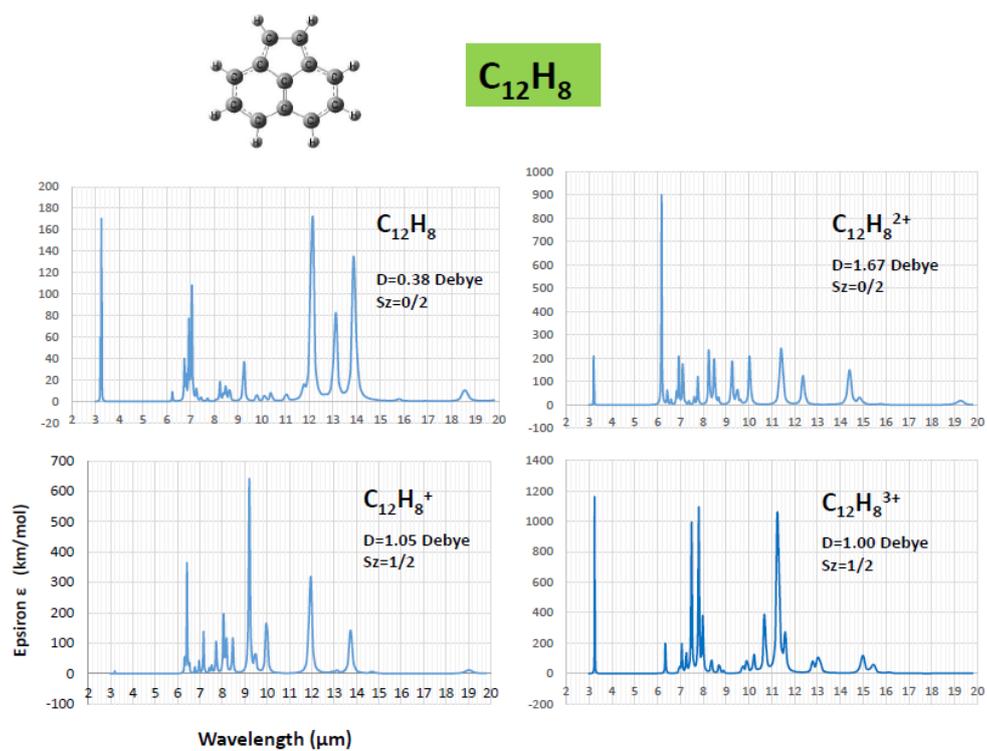

Figure 9, Calculated IR spectra of void induced 1H-phenalene $C_{12}H_8$.

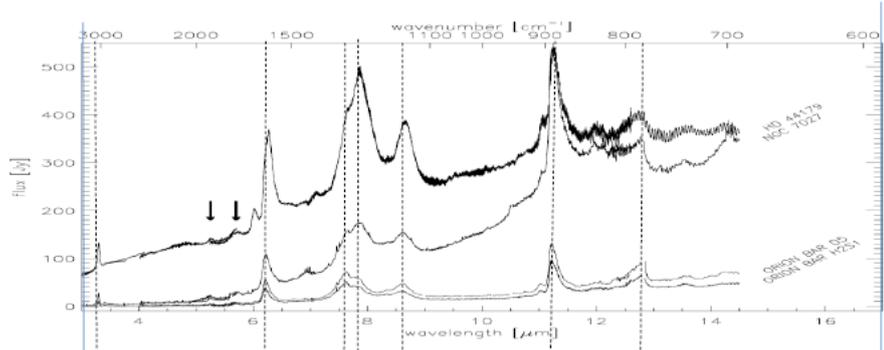
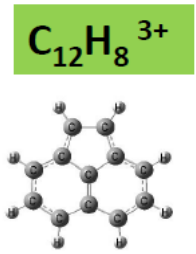
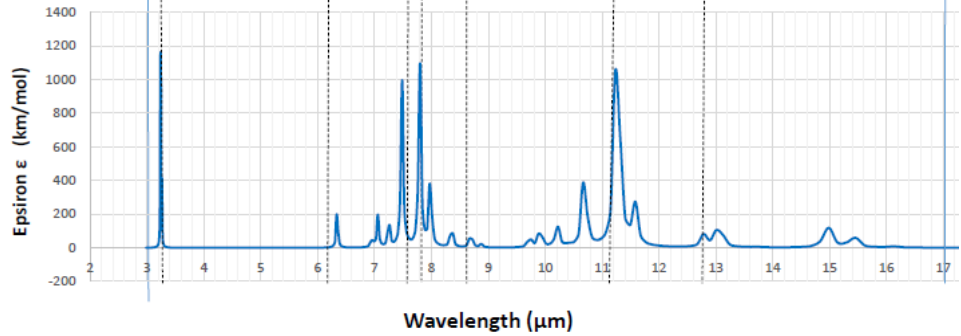

Figure 10, Comparison of calculated IR peaks of $C_{12}H_8^{3+}$ with astronomically observed spectra.

Table 4, Harmonic infrared wavelength and vibrational mode analysis of $C_8H_7^+$.

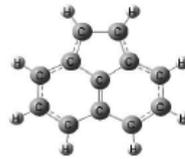
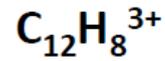

IR spectrum of $C_{12}H_8^{3+}$

| Well observed wavelength (micron) | Mode number | wavelength (micron) | epsiron (km/mol) | vibrational mode |
|---|---|---|---|---|
| | 1 | 72.8 | 0.7 | |
| | 2 | 63.9 | 0.0 | |
| | 3 | 62.3 | 33.1 | |
| | 4 | 39.2 | 0.0 | |
| | 5 | 30.4 | 8.1 | |
| | 6 | 29.7 | 0.0 | |
| | 7 | 26.6 | 5.2 | |
| | 8 | 25.0 | 85.7 | C-C in plane stretching |
| | 9 | 24.0 | 0.3 | |
| | 10 | 22.2 | 0.2 | |
| | 11 | 20.1 | 0.8 | |
| | 12 | 19.0 | 0.0 | |
| | 13 | 16.1 | 1.6 | |
| | 14 | 15.4 | 18.3 | |
| 14.3 | 15 | 15.0 | 35.2 | C-C and C-H out of plane bending |
| | 16 | 13.8 | 1.0 | |
| | 17 | 13.1 | 0.0 | |
| 13.5 | 18 | 13.1 | 42.2 | C-H out of plane bending |
| 12.7 | 19 | 12.8 | 21.0 | |
| | 20 | 11.6 | 69.8 | C-H out of plane bending |
| 11.2 | 21 | 11.3 | 420.2 | C-C stretching, C-H inplane bending |
| | 22 | 11.0 | 0.0 | |
| | 23 | 10.7 | 133.9 | C-C stretching, C-H inplane bending |
| | 24 | 10.4 | 3.5 | |
| | 25 | 10.3 | 0.0 | |

| Well observed wavelength (micron) | Mode number | wavelength (micron) | epsiron (km/mol) | vibrational mode |
|---|---|---|---|---|
| | 26 | 10.2 | 37.1 | |
| | 27 | 10.1 | 0.7 | |
| | 28 | 10.1 | 0.0 | |
| | 29 | 9.9 | 32.9 | C-C stretching, C-H inplane bending |
| | 30 | 9.7 | 17.7 | C-C stretching, C-H inplane bending |
| | 31 | 8.9 | 5.4 | |
| | 32 | 8.8 | 0.0 | |
| 8.6 | 33 | 8.7 | 25.0 | C-C stretching, C-H inplane bending |
| | 34 | 8.4 | 2.8 | |
| | 35 | 8.4 | 33.6 | C-C stretching, C-H inplane bending |
| | 36 | 8.0 | 127.4 | C-C stretching, C-H inplane bending |
| 7.8 | 37 | 7.8 | 327.5 | C-C stretching, C-H inplane bending |
| | 38 | 7.6 | 4.6 | |
| 7.6 | 39 | 7.5 | 287.9 | C-H inplane vibration |
| | 40 | 7.4 | 1.2 | |
| | 41 | 7.3 | 23.6 | |
| | 42 | 7.3 | 26.7 | |
| | 43 | 7.1 | 55.0 | C-C stretching, C-H inplane bending |
| | 44 | 7.0 | 7.8 | |
| | 45 | 6.9 | 11.3 | |
| 6.2 | 46 | 6.3 | 72.3 | C-C stretching, C-H in plane bending |
| | 47 | 3.3 | 1.4 | |
| | 48 | 3.3 | 101.7 | C-H stretching at hexagon |
| | 49 | 3.2 | 38.7 | |
| | 50 | 3.2 | 48.5 | |
| | 51 | 3.2 | 0.6 | |
| 3.3 | 52 | 3.2 | 217.8 | C-H stretching at pentagon |
| | 53 | 3.2 | 255.0 | C-H stretching at hexagon |
| | 54 | 3.2 | 0.7 | |

## 6, SUMMARY AND DISCUSSION

Judgement of coincidence with calculated IR spectrum compared with astronomically well observed one were summarized in Table 5. In an each column, calculated molecule with different charge number from 0 to +3 are added by colored number. Black marked charge number case show molecule having no good coincidence with observed spectrum, blue marked case fair, and blue marked case good. Judgement "Fair (green number)" means that among major IR peaks more than 50% peaks match with well observed wavelength within +/- 0.2μm, also "Good (blue number)" means overcome 75%. For example, $C_9H_7^+$ has eight major peaks as illustrated in Figure 6 and 7, among them six peaks coincide, two peaks does not, which brings judgement to be "fair". Exceptional case was $C_5H_5^+$, which show qualitatively "fair" behavior.

Comparing lines in Table 5, PAH (upper line) does not show any good IR feature even including cation cases. Whereas, some void induced PAH with cation cases (lower line) resulted "fair" and "good" judgement. In previous papers (Ota 2014, 2015a), void induced di-cation coronene $C_{23}H_{12}^{2+}$ were also analysed to be "good". In order to obtain fair or good IR spectrum, we can imagine necessarily two conditions, that is, the first one is asymmetrical molecular configuration having intrinsic electric dipole moment, and the second is ionization to bring cation molecule. Those are imagined to be made by high energy proton and photon attack on PAH's.

In interstellar space, there may be many symmetrical PAH molecules having no intrinsic dipole moment. In such case, we could not recognize them through observation of molecular emission spectrum. Whereas in case of asymmetrical and ionized molecules, there may be some capability to show strong emission spectra as like well astronomically observed one.

Table 5, Summary of calculated IR spectrum compared with astronomically well observed one. Under each column, calculated case for charge number from 0 to +3 are added. Black marked charge number case show no good coincidence with observed spectrum, blue marked case fairly good, and blue marked case good.

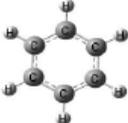

# 7, CONCLUSION

In order to clarify source molecules to show astronomically well observed ubiquitous infrared (IR) spectrum, simple polycyclic aromatic hydrocarbon molecules (PAH's) were analyzed by using the density functional theory (DFT). Starting molecules were benzene $C_6H_6$, naphthalene $C_{10}H_8$ and 1H-phenalene $C_{13}H_9$. In interstellar space, those molecules will be attacked by high energy photon and proton, which may bring cationic molecules like $C_6H_6^{2+}$, $C_{10}H_8^{3+}$ etc. also bring CH lacked molecules like $C_5H_5^{2+}$, $C_9H_7^{3+}$ etc. IR spectra of those molecules were analyzed based on DFT based Gaussian program. Results suggested that symmetrical configuration molecules as like benzene, naphthalene, 1H-phenalene and those cation ( +, 2+, and 3+) show little resemblance with observed IR. Contrast to such symmetrical molecules, several cases among void induced cationic molecules with asymmetrical configuration show fair and good IR behavior. One typical example was void induced 1H phenalene $C_{12}H_8^{3+}$. Calculated harmonic IR wavelength were 3.2, 6.3, 7.5, 7.8, 8.7, 11.3, and 12.8μm, which were compared with astronomically well observed wavelength of 3.3, 6.2, 7.6, 7.8, 8.6, 11.2, and 12.7μm. It was amazing agreement. Also, some cases like $C_5H_5^+$, $C_9H_7^+$, $C_9H_7^{2+}$, $C_9H_7^{3+}$, and $C_{12}H_8^{2+}$ show fair coincidence. Such results suggested that in case of asymmetrical and ionized cation molecules, there may be some capability to show strong emission spectrum similar with astronomically well observed one.

## ACKNOWLEDGEMENT

I would like to say great thanks to Dr. Christiaan Boersma, NASA Ames Research Center, to permit me to refer a figure (Boersma et al. 2009), also thanks his kind and useful suggestions.